\documentclass[a4paper, 11pt]{article}

\usepackage{a4wide}
\usepackage{amsmath}
\usepackage{latexsym}
\usepackage{graphicx}
\usepackage{color}
\usepackage[pdfstartview=FitH]{hyperref}
\usepackage{verbatim}
\usepackage{authblk}
\usepackage[subtle]{savetrees}

\title{A Problem in Human Dynamics: \\ Modelling the Population Density of a Social Space}

\author{Mark Levene mark@dcs.bbk.ac.uk}
\author{Trevor Fenner trevor@dcs.bbk.ac.uk}
\affil{Department of Computer Science and Information Systems, \authorcr Birkbeck, University of London, London WC1E 7HX, U.K.}

\date{}

\begin{document}

\maketitle

\begin{abstract}

Human dynamics and sociophysics suggest statistical models that may explain and provide us with better insight into social phenomena. Here we tackle the problem of determining the distribution of the population density of a social space over time by modelling the dynamics of agents entering and exiting the space as a birth-death process. We show that, for a simple agent-based model in which the probabilities of entering and exiting the space depends on the number of agents currently present in the space, the population density of the space follows a gamma distribution.
We also provide empirical evidence supporting the validity of the model by applying it to a data set of occupancy traces of a common space in an office building.

\end{abstract}


\section{Introduction}
\label{sec:int}

Recent interest in complex social systems, including social networks such as the world-wide-web, messaging and mobile phone networks \cite{ESTR11,NEWM18}, has led researchers to investigate the processes that could explain the dynamics of human behaviour within these networks.
The study of {\em human dynamics} \cite{FENN18a,YUAN18,LEVE19a} is not limited to the study of behaviour in communication networks, and has a broader remit similar to the aims of {\em sociophysics} \cite{GALA08,SEN14} (also formerly called {\em social physics}), which uses concepts and methods from statistical physics to investigate social phenomena such as opinion formation, voting behaviour and crowd dynamics \cite{CAST09}.
Social physics has a long history going back to the polymath Quetelet in the 19th century, who applied statistical laws to the study of human characteristics. For example, he derived the {\em body mass index} from the observation that body weight is approximately proportional to the square of body height \cite{EKNO08}. The foundations of 20th century social physics may be attributed to Stewart \cite{STEW50}, who introduced the concept of demographic gravitation, which applies gravitational potential theory to the geographic distribution of populations.
Of particular interest to human dynamics and social physics are {\em agent-based models} \cite{HAMI16,MACA16,NAMA16}, which overcome theoretical complexity by providing a computational mechanism for the simulation of social dynamics. Simulation of agent-based models can demonstrate the emergence of complex social behavioural patterns at the macro level resulting from simple interactions of individual agents at the micro level of the social system \cite{CAST09,SORN14}.

\smallskip

Monitoring and predicting occupancy in spaces is a well-known problem \cite{SAHA19}, which has implications, for example,
for optimising space usage \cite{LIU17}, detecting crowded spaces \cite{DEPA15}, recognising social interaction \cite{CRIV18}, and managing energy consumption \cite{SARA19}. Moreover, with the advent of wireless technologies such as WiFi, there is enhanced capability to scale up the process of monitoring human presence in environments such as universities \cite{ELDA18,MOHO18}. We provide additional detail in Section~\ref{sec:review}, where we give a brief overview of occupancy models in the context of building performance simulation \cite{HENS19}.

\smallskip

Viewing the problem of estimating occupancy presence as one of human dynamics, we tackle the related issue of determining the distribution of the population density of a {\em social space} (or simply a {\em space}) over time. We use the term social space to emphasise that the space allows interaction between its occupants,
as opposed to single-occupancy spaces such as single person offices. Examples of social spaces are: commercial buildings, open-plan office spaces, seminar rooms, restaurants, or even open-air spaces such as beaches that are spatially well-defined and have a limited capacity.
We model the dynamics of agents entering and exiting the social space as a {\em birth-death process} \cite{ALLE10,RENS11}; see Section~\ref{sec:sde}. In the basic model described in Section~\ref{sec:model}, agents enter or exit the space with probabilities that depend on the current population in the space; we show that, for this model, the population density of the space follows a gamma distribution \cite{JOHN94c}. To validate the basic model, simulation results are presented in Section~\ref{sec:simul}. We provide two extensions of the basic model: the first, in Section~\ref{sec:emigration}, allows agents to exit the space for reasons other than the level of occupancy, and the second, in Section~\ref{sec:agents}, allows for multiple types of agents. In Section~\ref{sec:data}, we provide a preliminary proof of concept for the model using a public data set \cite{LIU17} of occupancy data from a commercial office space collected over a period of nine months. Finally, in Section~\ref{sec:conc}, we give our concluding remarks.

\section{A brief review of occupancy models}
\label{sec:review}

Before developing our model in the following sections, we briefly review occupancy models used in {\em building performance simulation} (BPS) \cite{HENS19}. One of the main goals of BPS is to model various aspects of a building, including occupant presence, weather conditions and energy usage, with the intention of optimising energy efficiency throughout the building. On the one hand, occupancy can be predicted from environmental variables \cite{CAND16} such as temperature and sound, some of which are sources of energy,
and on the other hand occupancy modelling can lead to better energy management \cite{ERIC14}.
As a model is a simplification of reality, it is important to validate the model through simulations of real-life scenarios, using the simulation results to improve the model, whilst keeping the model as simple as possible.

\smallskip

We will focus on only one aspect of BPS, namely that of modelling the occupancy patterns in a building \cite{DOCA19}. Here  we are particularly interested in {\em social spaces}, where occupant interaction takes place, rather than single-occupancy spaces, where the concern is whether the occupant is present or absent from the space. Single-occupancy spaces can be modelled as Bernoulli random variables \cite{JOHN05a}, whereas modelling occupancy presence in social spaces, such as open-plan office spaces, involves more general statistical distributions of the population density of occupants. In general terms, when individuals are independent agents, Bernoulli random variables can be used to gauge the occupancy status of the building or a space therein \cite{YAN15}, for example, monitoring whether the space is empty or not.

\smallskip

Stochastic models of occupancy \cite{CHEN15} capture the stochastic nature of occupant behaviour in spaces and can be simulated in order to generate occupancy patterns \cite{FENG15}. Page et al. \cite{PAGE08} suggested a stochastic model for BPS, based on inhomogeneous Markov chains \cite{GARG09,SENE14}, where each occupant is modelled individually, assuming that behaviours of occupants are independent of each other. The inhomogeneous Markov chain allows Page et al. to model recurring daily occupancy  patterns, as well as long absences from a space and movement between different zones in a space. Chen et al. \cite{CHEN15} extend \cite{PAGE08} by modelling multiple occupants concurrently within an inhomogeneous Markov chain framework.

\smallskip

Occupancy models measure various quantities, key properties being: occupancy levels, presence and arrival/departure times.
Haldi \cite{HALD13} proposed a probabilistic model, based on generalised linear mixed models \cite{FOX16}, for predicting occupants' interactions with building components, in particular, actions on window openings and shading devices. The method allows for the representation of the diversity of individual behavioural profiles with a statistical distribution that describes individual interactions with the environment, allowing the inference of occupancy events such as arrival or departure. Feng  et al. \cite{FENG15} differentiate between four levels of occupancy, which vary in time: (i) the numbers of occupants in a building, (ii) the occupancy status of a space, (iii) the number of occupants in a space, and (iv) the space location of an occupant. They integrate levels (ii), (iii) and (iv) into a software simulation tool to capture these occupancy levels. Mahdavi and Tahmasebi \cite{MAHD15} empirically tested the predictions of occupancy profiles for several occupancy models and found the predictions to be rather low. However, a model returning an aggregated profile of presence probability performed best, where presence was predicted when the probability was above a specified threshold.

\smallskip

Agent-based models simulate human behaviours with software agents that encode the behaviours of the systems being studied \cite{HAMI16}. Computer simulation of the resulting agent-based model can then be used to generate behavioural traces of occupants, which can be aggregated to form occupancy patterns of the space being modelled.
Liao et al. \cite{LIAO12} made use of agent-based modelling to extend the stochastic model of Page et al. \cite{PAGE08} to an arbitrary number of occupants and zones within a building. They also proposed a graphical model \cite{KOLL09} with reduced complexity, which is shown to have comparable predictive accuracy to the agent-based model in describing the mean occupancy in the building. Chen et al. \cite{CHEN18} present an agent-based Occupancy Simulator that captures individual profiles of stochastic behaviours. The simulator is able to perform a detailed stochastic simulation of occupants' presence and movement within a building by integrating several existing stochastic occupancy models, in particular, those of Wang et al. \cite{WANG11} and Reinhart \cite{REIN04}. Luo et al. \cite{LUO17} provided an evaluation of the Occupancy Simulator for a real-world occupancy data set, and showed that the simulator can accurately reproduce a wide variety of occupancy patterns.

\smallskip

The model we present herein is agent-based and is formalised as a birth-death stochastic process, where birth corresponds to arrival and death to departure; see Sections \ref{sec:sde} and \ref{sec:model}. The only occupancy model we are aware of that borrows directly from birth-death processes is that of Jia and Spanos \cite{JIA17}, where occupancy is modelled as an infinite-server with time-varying arrival and departure rates; we note that many queue types, such as the infinite-server queue used in \cite{JIA17}, can be modelled as birth-death processes \cite{SHOR18}. Here we concentrate on providing a statistical estimate of the number of occupants in a social space, with the aid of an agent-based stochastic occupancy model that generates a distribution capturing the density of the population in the space. Our formalism allows us to derive the equilibrium density in the space when the occupancy level is in a steady state \cite{ALLE10,RENS11}. Thus our model complements previous stochastic models in BPS and could potentially be integrated into an existing agent-based simulation tool.

\smallskip

More recently, several machine learning algorithms have been developed for occupancy detection and prediction \cite{SAHA19}. In particular, applying neural networks \cite{AGGR18} seems promising, as in \cite{CHEN17}, where deep neural networks are deployed to identify the level of occupancy as zero, low, medium or high.

\smallskip

Finally, we note that in terms of the hardware employed for occupancy detection, as well as commonly used sensor technologies \cite{CRIV18,SARA19},
wireless technologies, such as WiFi, can also be used \cite{DEPA15,ELDA18,MOHO18}.

\section{Stochastic Differential Equations and Birth-Death Processes}
\label{sec:sde}

Here we present some background to the methods we use to tackle the occupancy density problem for social spaces.
In particular, we present the {\em Ornstein-Uhlenbeck} process; this takes the form of a mean-reverting {\em stochastic differential equation} (SDE)
\cite{GULI12,HIRS14} that leads to stationary distributions \cite{BIBB05}, including the class of Pearson diffusions \cite{FORM08}.
We also give the necessary background on birth-death stochastic processes and their description by SDEs \cite{ALLE10,RENS11}.

\smallskip

An {\em Ornstein-Uhlenbeck} process is described by an SDE that takes the form
\begin{equation}\label{eq:sde}
{\mathrm d} X_t = - \theta\left( X_t - m \right) {\mathrm d} t + \sigma(X_t) {\mathrm d} W_t,
\end{equation}
where $X_t$ is a random variable and $t$ is a positive real parameter that denotes time. The positive constant $\theta$ is the {\em rate} parameter,
$m$ is the mean of the underlying stochastic process, $\sigma(\cdot)$ is the {\em diffusion} function, and $W_t$ is a Wiener process (also known as Brownian motion); the term $- \theta\left(X_t - m\right)$ is known as the {\em drift} of the process.

\smallskip

The equilibrium density $f(x)$ of the SDE (\ref{eq:sde}) is given by
\begin{equation}\label{eq:equilibrium}
f(x) = \frac{\kappa_0}{\sigma^2(x)} \exp\left( -2 \theta \int_0^x \frac{y - m}{\sigma^2(y)} {\mathrm d}y \right),
\end{equation}
for some positive constant $\kappa_0$ \cite{BIBB05}.

\smallskip

For a birth-death process, let $X_t$ represent the population size at time $t$, $\lambda(X_t)$ denote the birth rate and $\mu(X_t)$ denote the death rate. Then, the SDE corresponding to the birth-death process \cite{ALLE10,GRAN15a}, is given by
\begin{equation}\label{eq:bd}
{\mathrm d} X_t = \Bigl( \lambda(X_t) - \mu(X_t) \Bigr) {\mathrm d} t + \sqrt{\lambda(X_t) + \mu(X_t)} {\mathrm d} W_t.
\end{equation}
\smallskip

\section{Model Assumptions and Solution}
\label{sec:model}

In the scenarios we investigate, we assume a single social space with fixed capacity.
Agents enter and exit the space according to the birth-death process described below.
Entering the space, i.e. arrival, corresponds to {\em birth} and exiting the space, i.e. departure, corresponds to {\em death}. In this and the following two sections, we assume that all agents are of the same type.

\smallskip

Let the capacity of the social space in question be $C$, and the population in the space at time $t$ be $X_t$. We make use of two positive parameters $\alpha$ and $\beta$, where $\beta \ge \alpha > 0$, corresponding respectively to the preferences that agents may have regarding entering and exiting the space. The birth rate, given in (\ref{eq:birth}) below, is proportional to the fraction of the capacity currently unfilled,
while the death rate, given in (\ref{eq:death}) below, is proportional to the fraction of the capacity currently filled.

\smallskip

In many cases, individual behaviour is governed by time, for example, when arriving at work in the morning and departing in the late afternoon. Indeed, time is one of the most important factors affecting arrival and departure rates. Goldstein et al. \cite{GOLD10}, however, argued that multiple factors affect occupants' behavioural patterns; examples of such factors are: the time of day, the occupancy pattern of the space, and the indoor and outdoor weather conditions \cite{ZHAN12a}.
Here we make a case for the occupancy level, specified as a fraction of the capacity of the space, to be considered as another important factor affecting occupants' behaviour. Now, assuming that people do not all arrive at work at exactly the same time, the rate of arrival goes down
as more people occupy the space. At some stage, when most of the people are at work, a steady state is attained, i.e. the population in the space stabilises; the tendency over time of the population density to reach an equilibrium is formally shown below in the derivation of the model.
Considering this example further, departures from the space, when people leave work, does not strictly follow our model, as peoples' tendency to leave work increases towards the end of the day, independently of the number of people in the space. However, we show in Section~\ref{sec:emigration} that it is possible to model departures from a social space for reasons other than the degree of crowdedness of the space by introducing an emigration constant into the birth-death model \cite{ALLE10,GRAN15a}; see Section~\ref{sec:data} for more discussion on this with respect to a real-world occupancy data set. Nevertheless, there are many natural scenarios in which the rate of leaving a space is proportional to its current occupancy. In general, it is fair to say that, when a space is crowded, people's tendency to leave the space is proportional to how crowded the space is; for example, consider a crowded bar or a public space such as a beach.
(We note that there is a symmetry between arrivals and departures, even though the rates may differ.)

\smallskip

Continuing the formalisation of the model, when the population in the space is $X$,
\begin{equation}\label{eq:birth}
\lambda(X) = \alpha \left( 1 - \frac{X}{C} \right),
\end{equation}
and
\begin{equation}\label{eq:death}
\mu(X) = \beta \left( \frac{X}{C} \right).
\end{equation}
\smallskip

We note that the above two functions are linear in $X$.

\smallskip

Subtracting (\ref{eq:death}) from (\ref{eq:birth}) gives
\begin{equation}\label{eq:bminusd}
\lambda(X) - \mu(X) = \alpha - \frac{X}{C} \left( \alpha + \beta \right),
\end{equation}
and adding them gives
\begin{equation}\label{eq:bplusd}
\lambda(X) + \mu(X) = \alpha + \frac{X}{C} \left( \beta - \alpha \right).
\end{equation}
\smallskip

Substituting these into (\ref{eq:bd}), we see that we need to solve (\ref{eq:sde}) where
\begin{equation}\label{eq:drift}
- \theta\left( X - m \right) = \alpha - \frac{X}{C} \left( \alpha + \beta \right) = - \left(\frac{\alpha+\beta}{C}\right) \Biggl( X - \alpha \left(\frac{C}{\alpha+\beta}\right)\Biggr),
\end{equation}
and
\begin{equation}\label{eq:diffusion}
\sigma^2(X) = \alpha + \frac{X}{C} \left( \beta - \alpha \right).
\end{equation}

Letting $\delta = \beta - \alpha$, we obtain
\begin{equation}\label{eq:theta-delta}
\theta = \frac{2 \alpha + \delta}{C},
\end{equation}
\begin{equation}\label{eq:theta-m}
m = \frac{\alpha C}{2 \alpha + \delta}
\end{equation}
and
\begin{equation}\label{eq:sigma}
\sigma^2(X)=  \frac{\delta X + \alpha C}{C}.
\end{equation}
\smallskip

Now, from (\ref{eq:equilibrium}), we deduce that the equilibrium density is given by
\begin{equation}\label{eq:eq-bd}
f(x)  = \left( \frac{\kappa_0 C}{\delta x + \alpha C} \right) \exp\left( - 2 \int_0^x \frac{\left( 2 \alpha + \delta \right) y - \alpha C}{\delta y + \alpha C} {\mathrm d}y \right).
\end{equation}
\smallskip

In the special case when $\delta = 0$, we have
\begin{equation}\label{eq:eq-bd-sc}
f(x)  = \kappa_1 \exp\Biggl( - 2 \int_0^x \left( \frac{2 y}{C} - 1 \right ) {\mathrm d}y \Biggr),
\end{equation}
for some positive constant $\kappa_1$.
On evaluating the integral, we obtain
\begin{equation}\label{eq:int-sc}
f(x)  = \kappa_1 \exp \Biggl( - 2 \left( \frac{x^2}{C} - x \right)  \Biggr).
\end{equation}
\smallskip

Rearranging (\ref{eq:int-sc}) gives
\begin{equation}\label{eq:int-norm}
f(x)  = \kappa_2 \exp \Biggl( -  \frac{\left( x - \frac{C}{2}\right)^2}{ 2 \left( \frac{C}{4} \right)} \Biggr),
\end{equation}
for some positive constant $\kappa_2$, i.e., a normal density function with mean  $\frac{C}{2}$ and variance $\frac{C}{4}$.

\smallskip

For the general case of (\ref{eq:eq-bd}), where $\delta > 0$, we obtain
\begin{equation}\label{eq:int0-gamma}
f(x)  = \kappa_3 \left( \delta x +  \alpha C \right)^{\frac{4 \alpha C (\alpha + \delta)}{\delta^2} -1} \exp \left( - \frac{ (4 \alpha + 2 \delta) x}{\delta} \right),
\end{equation}
for some positive constant $\kappa_3$.
This solution is only valid if $\delta x + \alpha C > 0$, otherwise the value of $f(x)$ is generally complex and is unbounded.
For this reason, we have assumed that $\beta \ge \alpha$ (i.e. the death parameter is greater than the birth parameter).

\medskip

A gamma distribution \cite{JOHN94c} has probability density function
\begin{equation}\label{eq:gamma}
\frac{\phi^\tau}{\Gamma(\tau)} (x - \nu)^{\tau-1}  \exp \Bigl(- \phi (x - \nu) \Bigr),
\end{equation}
where $\Gamma$ is the gamma function \cite[6.1]{ABRA72}; $\tau > 0$ is the {\em shape} of the distribution, $\phi > 0$ is the {\em rate}, and $\nu$ is the {\em location} parameter.
The mean of the gamma distribution is $\frac{\tau}{\phi} + \nu$ and its variance is $\frac{\tau}{\phi^2}$.
For large shape parameter $\tau$, the gamma distribution with shape $\tau$ is close to a normal distribution with the same mean and variance
\cite[Section 10.5]{DASG10}.

\smallskip

If can be seen that $f(x)$ in (\ref{eq:int0-gamma}) is a gamma density function, and it is straightforward to show that the mean and variance tend to $\frac{C}{2}$ and $\frac{C}{4}$, respectively, as $\delta$ tends to $0$.

\smallskip

For simplicity in the computations that follow, we assume that $\delta = 1$; we also set $\rho = \alpha C$. In this case, equation~(\ref{eq:int0-gamma}) reduces to
\begin{equation}\label{eq:int1-gamma}
f(x)  = \kappa_4 \left( x + \rho \right)^{4 \alpha (\rho +C) -1} \exp \Bigl( - (4 \alpha +2) (x + \rho) \Bigr),
\end{equation}
for some positive constant $\kappa_4$.
In this case, the parameters in (\ref{eq:gamma}) are $\tau = 4 \alpha (\rho +C)$, $\phi = 4 \alpha + 2$, and $\nu = -\rho$.
From these parameters, we derive the mean and variance of the model as, respectively,
\begin{equation}\label{eq:int1-mean-std}
\mu_{_{\cal M}} = \frac{\rho}{2 \alpha +1} \ \ {\rm and} \ \ \sigma^2_{_{\cal M}} = \frac{(\alpha +1) \rho}{(2 \alpha +1)^2}.
\end{equation}
\smallskip


We observe that to formally deal with the lower and upper boundaries, 0 and $C$, we could incorporate reflecting barriers \cite{WARD03b,HARR13}, resulting in a truncated gamma density as the equilibrium distribution. In the special case when $\alpha = \beta$, or for large $\tau$,
the resulting distribution is, or approaches, a truncated normal density \cite{CHA13,THOM18}; see \cite{WARD03a,PEND15} for a similar result in the context of single server queues with abandonment.

\section{Simulations of the Basic Model}
\label{sec:simul}

We can simulate the stochastic process corresponding to (\ref{eq:bd}) using the {\em Euler-Maruyama method} \cite{SAUE13}, which is a general computational method for obtaining approximate numerical solutions to SDEs. We will also make use of the {\em Jensen-Shannon divergence} (JSD) \cite{ENDR03} as a goodness-of-fit measure \cite{LEVE18}. All computations were carried out using the Matlab software package.

\smallskip

The results of a typical simulation run of the process, using the Euler-Maruyama method to generate the sequence $X_1, X_2, \ldots, X_T$ for $T = 10^6$, are summarised in Table~\ref{table:sim}; several repeated simulations produced very similar results.
The parameters used were $C = 100$, $\delta = 1$ and $\alpha$ ranging from $0.05$ to $0.95$ in steps of $0.05$.
For completeness purposes, the bottom row in the table shows the results for $\delta = 0$, i.e., when $\alpha = \beta = 1$.
We fitted first normal and then gamma distributions to the sequence of values of $X_t$ obtained from the simulations for each value of $\alpha$, using the maximum likelihood method. The means and standard deviations of the fitted distributions are shown in Table~\ref{table:sim} as
$\mu_{_{\cal N}}$ and $\sigma_{_{\cal N}}$ for the normal distribution ($\cal N$), and $\mu_{_\Gamma}$ and $\sigma_{_\Gamma}$ for the gamma distribution ($\Gamma$). The JSD values show a very good fit of both the computed normal and gamma distributions to the sequence $(X_t)$ generated by the simulations. As expected, the fit to the normal distribution gets better as $\alpha$ gets closer to $\beta$, since $\tau$ increases more than linearly with $\alpha$. The last row, of course, shows an exceptionally good fit, as predicted by (\ref{eq:int-norm}).
The last two columns, headed $\mu_{_{\cal M}}$ and $\sigma_{_{\cal M}}$, show the mean and standard deviation of the distribution predicted by the model ($\cal M$) as given in (\ref{eq:int1-mean-std}), or  (\ref{eq:int-norm}) for the special case when $\alpha = \beta$. It can be seen from Table~\ref{table:sim} that these values are close to those of the fitted normal and gamma distributions.

\begin{table}[ht]
    \begin{center}
    \begin{tabular}{lrrrrrrrr}
    $\alpha$ & $\mu_{_{\cal N}}$ & $\sigma_{_{\cal N}}$ & JSD${}_{_{\cal N}}$ & $\mu_{_\Gamma}$ & $\sigma_{_\Gamma}$ & JSD${}_{_\Gamma}$ & $\mu_{_{\cal M}}$ & $\sigma_{_{\cal M}}$ \\ \hline
    0.05 & 4.5879  & 2.0694 & 0.0227 & 4.5879  & 2.3684 & 0.0337 & 4.5455  & 2.0830 \\
    0.10 & 8.3317  & 2.7429 & 0.0128 & 8.3317  & 2.9151 & 0.0208 & 8.3333  & 2.7639 \\
    0.15 & 11.5029 & 3.1926 & 0.0116 & 11.5029 & 3.3046 & 0.0153 & 11.5385 & 3.1949 \\
    0.20 & 14.3241 & 3.5009 & 0.0117 & 14.3241 & 3.5838 & 0.0118 & 14.2857 & 3.4993 \\
    0.25 & 16.6922 & 3.7051 & 0.0076 & 16.6922 & 3.7912 & 0.0131 & 16.6667 & 3.7268 \\
    0.30 & 18.7476 & 3.9238 & 0.0074 & 18.7476 & 4.0022 & 0.0122 & 18.7500 & 3.9031 \\
    0.35 & 20.5949 & 4.1280 & 0.0064 & 20.5949 & 4.2102 & 0.0120 & 20.5882 & 4.0434 \\
    0.40 & 22.1518 & 4.1932 & 0.0073 & 22.1518 & 4.2562 & 0.0102 & 22.2222 & 4.1574 \\
    0.45 & 23.7242 & 4.2797 & 0.0048 & 23.7242 & 4.3498 & 0.0117 & 23.6842 & 4.2514 \\
    0.50 & 25.0170 & 4.3221 & 0.0053 & 25.0170 & 4.3837 & 0.0105 & 25.0000 & 4.3301 \\
    0.55 & 26.1306 & 4.3864 & 0.0059 & 26.1306 & 4.4468 & 0.0096 & 26.1905 & 4.3967 \\
    0.60 & 27.2279 & 4.4744 & 0.0041 & 27.2279 & 4.5391 & 0.0110 & 27.2727 & 4.4536 \\
    0.65 & 28.2400 & 4.5495 & 0.0045 & 28.2400 & 4.6140 & 0.0101 & 28.2609 & 4.5027 \\
    0.70 & 29.1026 & 4.5791 & 0.0045 & 29.1026 & 4.6365 & 0.0098 & 29.1667 & 4.5453 \\
    0.75 & 29.9590 & 4.5850 & 0.0041 & 29.9590 & 4.6447 & 0.0098 & 30.0000 & 4.5826 \\
    0.80 & 30.7462 & 4.6391 & 0.0043 & 30.7462 & 4.6940 & 0.0094 & 30.7692 & 4.6154 \\
    0.85 & 31.4944 & 4.6699 & 0.0032 & 31.4944 & 4.7261 & 0.0102 & 31.4815 & 4.6444 \\
    0.90 & 32.2133 & 4.7254 & 0.0035 & 32.2133 & 4.7814 & 0.0097 & 32.1429 & 4.6702 \\
    0.95 & 32.8005 & 4.7450 & 0.0029 & 32.8005 & 4.8007 & 0.0102 & 32.7586 & 4.6933 \\
    {\em 1.00} & {\em 50.0182} & {\em 4.9842} & {\em 0.0010} & {\em 50.0182} & {\em 5.0279} & {\em 0.0095} & {\em 50.0000} & {\em 5.0000} \\
    \end{tabular}
    \caption{\label{table:sim} Simulation results for a typical run of the Euler-Maruyama method.}
    \end{center}
\end{table}
\smallskip

\section{An Extension to Allow Emigration}
\label{sec:emigration}

Here we extend the death rate to allow {\em emigration} \cite{ALLE10,GRAN15a},  in order to model the situation when an agent may leave the social space for reasons other than the space being crowded; for example, if the agent has fulfilled a certain task or a certain time has elapsed.

\smallskip

We modify (\ref{eq:death}) to
\begin{equation}\label{eq:death-em}
\mu(X) = \beta \left( \frac{X}{C} \right) + \gamma,
\end{equation}
where $\gamma$ is some non-negative emigration constant.

\smallskip

Following the derivation of (\ref{eq:eq-bd}) in Section~\ref{sec:model}, but now using (\ref{eq:death-em}) instead of (\ref{eq:death}), we get
\begin{equation}\label{eq:eq-bd-em}
f(x)  = \left( \frac{\kappa_0 C}{\delta x + (\alpha + \gamma) C} \right) \exp\left( - 2 \int_0^x \frac{\left( 2 \alpha + \delta \right) y - (\alpha - \gamma) C}{\delta y + (\alpha + \gamma) C} {\mathrm d}y \right),
\end{equation}
for some positive constant $\kappa_0$.

\smallskip

Corresponding to (\ref{eq:int0-gamma}), on evaluating the integral in (\ref{eq:eq-bd-em}), we obtain
\begin{equation}\label{eq:int0-gamma-em}
f(x)  = \kappa_5 \left( \delta x +  (\alpha + \gamma) C \right)^{\frac{4 \alpha C (\alpha + \gamma + \delta)}{\delta^2}-1} \exp \left( - \frac{(4 \alpha + 2 \delta) x}{\delta} \right),
\end{equation}
for some positive constant $\kappa_5$.

\smallskip

If we again assume that $\delta = 1$ and now set $\tilde{\rho} = (\alpha + \gamma) C$ then, corresponding to (\ref{eq:int1-gamma}), this reduces to \begin{equation}\label{eq:int1-gamma-em}
f(x)  = \kappa_6 \left( x +  \tilde{\rho} \right)^{4 \alpha (\tilde{\rho} +C)-1} \exp \Bigl( - (4 \alpha +2) (x + \tilde{\rho}) \Bigr),
\end{equation}
for some positive constant $\kappa_6$; thus $f(x)$ is a gamma density function with shape $\tau = 4 \alpha (\tilde{\rho} +C)$, rate $\phi = 4 \alpha + 2$, and location $\nu = -\tilde{\rho}$.

\smallskip

We note that a similar extension to the birth rate, where an agent may enter the space for reasons other than it being sparsely populated,
would allow us to incorporate {\em immigration} into the model  \cite{ALLE10,GRAN15a}.

\section{An Extension to Allow Multiple Types of Agents}
\label{sec:agents}

We now provide a restricted extension of the model presented in Section~\ref{sec:model} to multiple types of agents.
We concentrate on a special solvable case for two or more types of agents, which reduces to the one-dimensional case solved in Section~\ref{sec:model}.
For simplicity, we assume that there are just two types of agents, and that in the social space there are $X_1$ agents of type one and $X_2$ of type two.

\smallskip

In order for the model to be solvable, we make the assumption that the birth and death rates depend only on $X$, the total number of agents in the space, where $X = X_1 + X_2$. The reasoning behind this assumption is that, when an agent, say of type one, is contemplating entering the space (i.e., a birth event), it does not need to consider how many of the $X$ agents already in the space are of type one, since all that concerns the agent is whether, from its point of view, the space has sufficient extra capacity; this depends solely on $X$ and not on its breakdown into $X_1$ and $X_2$.
Similarly, when an agent of type one is contemplating exiting the space (i.e., a death event), it only needs to consider the overall occupancy of the space. That is, the only consideration of an agent is the objective perception of how crowded the space is.

\smallskip

For two agent types having $\lambda_1(\cdot)$ and $\lambda_2(\cdot)$ as their birth rates, we have
\begin{equation}\label{eq:birth2}
\lambda(X) = \lambda_1(X) + \lambda_2(X) = \left( \alpha_1 + \alpha_2 \right) \left( 1 - \frac{X}{C} \right),
\end{equation}
where $\alpha_1$ and $\alpha_2$ are the preferences of the two agents types regarding entering the space.
Similarly, if the death rates are $\mu_1(\cdot)$ and $\mu_2(\cdot)$, we have
\begin{equation}\label{eq:death2}
\mu(X) = \mu_1(X) + \mu_2(X) = \left( \beta_1 + \beta_2 \right) \left( \frac{X}{C} \right),
\end{equation}
where $\beta_1$ and $\beta_2$ are the preferences of the two agents types regarding exiting the space.
The solution of the space density problem is now given by (\ref{eq:int0-gamma}) with $\alpha = \alpha_1 + \alpha_2$ and $\beta = \beta_1 + \beta_2$.

\section{An Application to Occupancy of Office Space}
\label{sec:data}

There is a growing literature on measuring occupancy and usage in buildings \cite{SAHA19}, as reviewed in Section~\ref{sec:review}.
Simulation of occupancy patterns via stochastic \cite{CHEN15,FENG15} and agent-based models \cite{LIAO12,LUO17} provides the foundation for developing tools that measure physical properties of a building, including energy performance and the interaction of occupants with the environment.
Occupancy, i.e. the distribution of the number of people in the building over time, is an important aspect of occupants' behaviour and can be modelled through simulation, thereby informing prospective optimisation of the building's performance.

\smallskip

\begin{figure}[htb]
\centering{\includegraphics[scale=0.75]{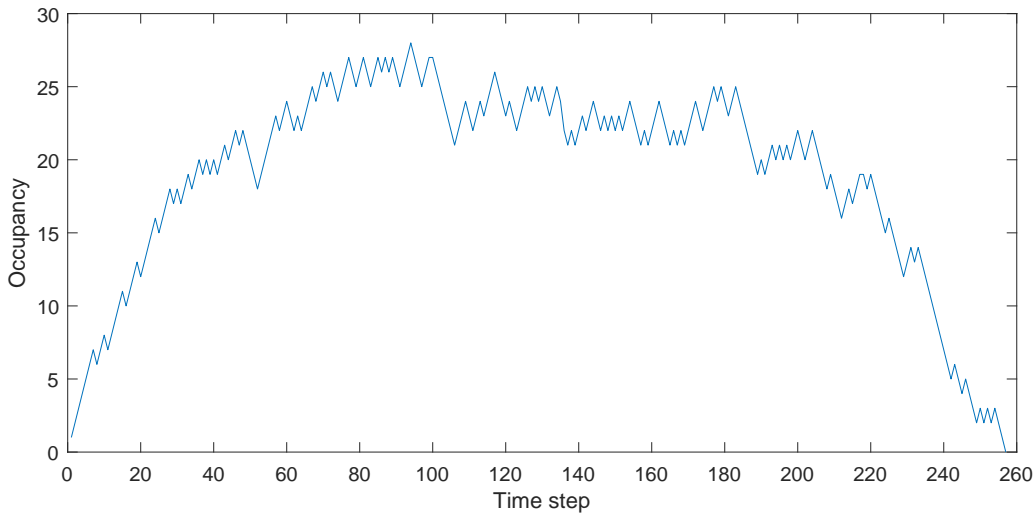}}
\caption{\label{figure:occ} Occupancy time series of the common office space on a typical weekday.}
\end{figure}

It is desirable to validate occupancy models on real-world data sets. Unfortunately, at the time of writing, there is a general lack of publicly available occupancy data sets that would allow us to test our model. However, Liu et al. \cite{LIU17}, who collected occupancy data from a commercial office space during 181 days over a period of nine months, have made their data set available, enabling us to present a preliminary proof of concept of our model. In particular, we make use of the occupancy patterns of the common office space derived from this data set, since our objective is to estimate the population density of the space.
In Figure~\ref{figure:occ}, we show a time series for the occupancy of this space during a typical weekday, where people start arriving at about 8:00 and the last person normally leaves before 20:00. Occupancy measurements were taken regularly every few minutes during the day to capture arrival and departure events as they occurred. In this context, it is interesting to mention that Ahn and Park \cite{AHN19} demonstrated that varying the sampling interval between 5 and 60 minutes does not significantly affect prediction of the number of occupants in a space. We further note that this data set only records net occupancy since, if both an arrival and a departure occurred within the interval of measurement, the net effect is that the number of occupants in the space remains the same. This can be viewed as a limitation of our case study, since the estimated birth and death rates will be underestimates of the true rates; we note that the smaller the interval of measurement the less likely this problem is to occur.

\smallskip

As can be seen in Figure~\ref{figure:occ}, a typical daily time series has three phases. During the {\em first phase}, people arrive at the office. During the {\em second phase}, most of the employees are already in the office and the number of arrivals and departures is fairly balanced -- for example, people take breaks or have meetings outside the office, but overall the number of occupants is relatively constant. Finally, during the {\em third phase}, people leave the office after a day's work.

\smallskip

Looking at the three phases from the perspective of our model, the first phase is generally modelled by an arrival process, since the space is initially empty. In the second, steady-state phase, we may consider the density of occupancy to be close to equilibrium, with fairly balanced numbers of arrivals and departures. The final third phase is dominated by departures from the space.

\smallskip

In the subsequent analysis, we have set the capacity $C$ to be 54 (which was the maximum occupancy over the 181 day period), and
we have assumed for simplicity that we have only one type of agent with no emigration and that $\alpha = \beta = 1$.
Recalling from the discussion at the end of Section~\ref{sec:model} that, in this case, the limiting equilibrium distribution is a truncated normal, it is reasonable to assume that, during the second phase, the occupancy is between 14 and 28, where 14 is the average minimum occupancy of the space over the 181 days period and 28 is the average maximum occupancy. In this case, the fitted normal distribution truncated below at 14 and above at 28 has a mean of $20.72$ and a variance of $10.80$; the JSD of the fitted normal distribution is $0.026$, which implies a good fit.
We note that the ratio of mean to variance is close to 2, as predicted by the model according to (\ref{eq:int-norm}).
Although the model predicts the mean to be $\frac{C}{2}$, i.e., 27, we observe that only on 10 days was the occupancy more than twice the predicted mean, which suggests that setting $C$ to be 54 is an overestimate vis-a-vis the model. A possible explanation for the low value of the mean is that on a number of days (for example, around the Christmas/New Year period) the occupancy is particularly low.

\smallskip

\begin{figure}[htb]
\centering{\includegraphics[scale=0.75]{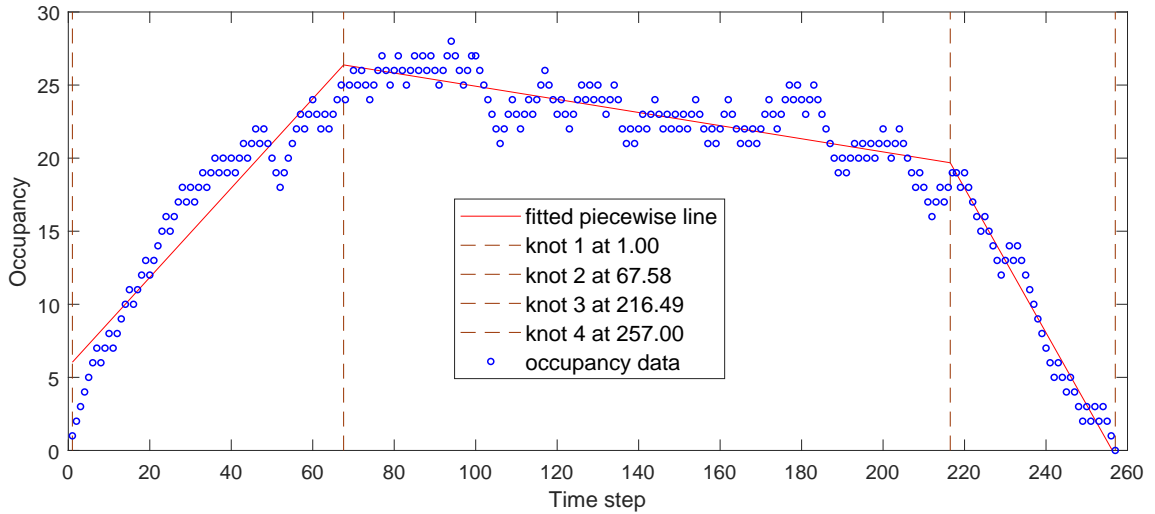}}
\caption{\label{figure:slm-occ} Occupancy time series output from SLM annotated with the fitted piecewise linear regression.}
\end{figure}

In order to differentiate the three phases, we employed the Matlab package, Shape Language Modelling ({\em SLM}) \cite{ERRI17}, which implements piecewise linear regression \cite{MARS01}. In particular, we stipulated four knots (also known as breakpoints) allowing for free placement of the two interior knots, which specify the beginning and end of the second phase. For example, for the typical day shown in Figure~\ref{figure:occ},
the output from SLM is depicted in Figure~\ref{figure:slm-occ}.
The first phase occurs between steps 1 (knot 1) and 67 (knot 2), the second phase occurs between steps 68 (knot 2) and 217 (knot 3), and the third phase occurs between steps 218 (knot 3) and 257 (knot 4), the final time step of the day.

\smallskip

To validate the model, for each {\em occupancy number} (i.e. number of occupants) from 1 to the maximum possible occupancy 54, we compute estimates from the data of the birth and death rates over the 181 days. In each case the data was smoothed with a moving average over a centred sliding window of length 25.
We then used linear regressions to fit the data to the model, according to (\ref{eq:birth}) and (\ref{eq:death}). Goodness of fit for the linear regressions was measured by the coefficient of determination $R^2$ \cite{MOTU95}, with values close to one indicating very good fits.
The arrival and departure rates were normalised to add up to one, since we have assumed that $\alpha = \beta = 1$; see (\ref{eq:bplusd}).

\begin{figure}[htb]
\centering{\includegraphics[scale=0.6]{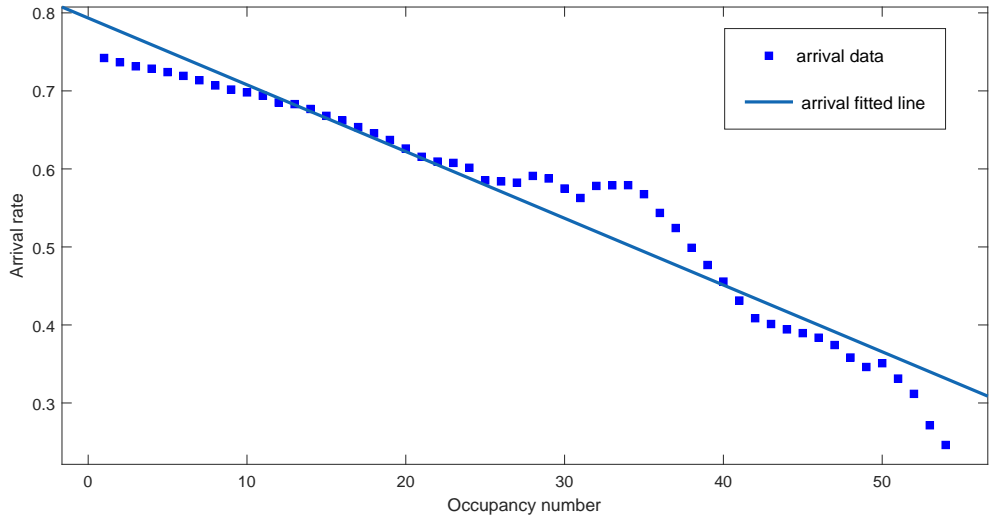}}
\caption{\label{figure:phase1} Linear regression of phase 1 arrival data.}
\end{figure}
\begin{figure}[htb]
\centering{\includegraphics[scale=0.6]{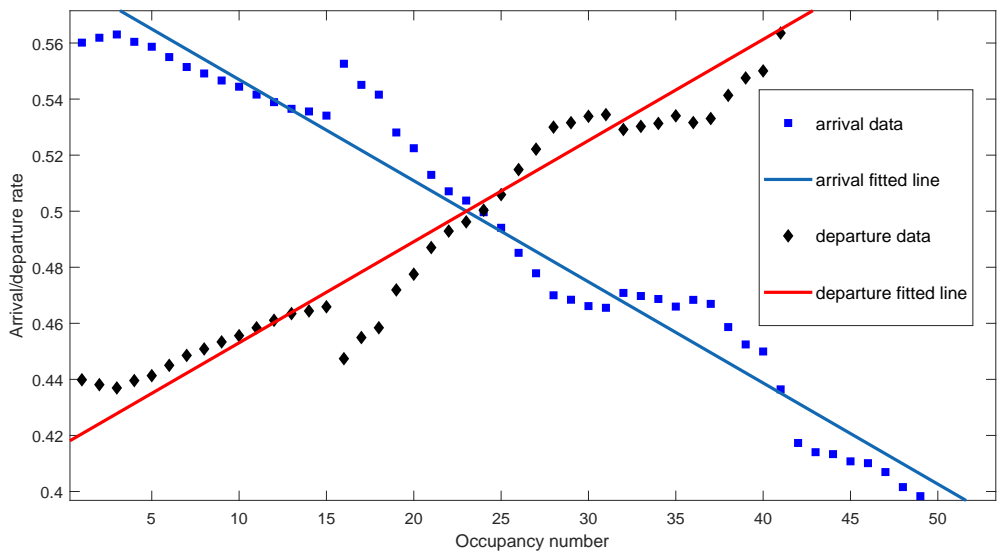}}
\caption{\label{figure:phase2} Linear regression of phase 2 arrival and departure data.}
\end{figure}
\begin{figure}[htb]
\centering{\includegraphics[scale=0.6]{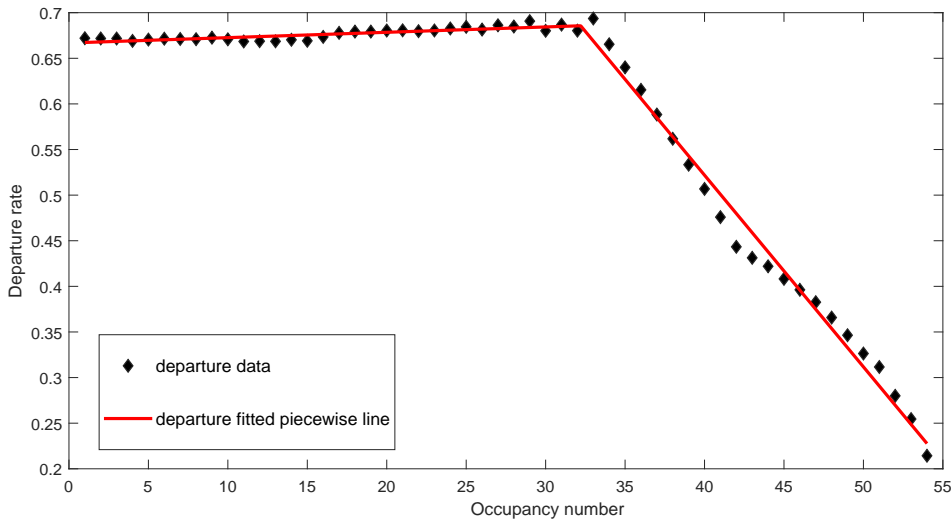}}
\caption{\label{figure:phase3} Piecewise linear regression of phase 3 departure data.}
\end{figure}
\smallskip

The computations were carried out as follows:
\renewcommand{\labelenumi}{(\roman{enumi})}
\begin{enumerate}
\item We compute just the empirical birth rate for the first phase, as we do not expect the death rate during the first phase to follow a linear trend with respect to occupancy; see Figure~\ref{figure:phase1} for the resulting linear regression, where $R^2 = 0.9417$. We note that arrival rates approximately follow a linear downwards trend, in accordance with the model as in (\ref{eq:birth}).

\item We compute the empirical birth and death rates for the second phase; see Figure~\ref{figure:phase2} for the resulting linear regressions, where both have $R^2 = 0.9681$. Since, in the second phase, both arrival and departure rates were taken into consideration in the normalisation, and thus for each occupancy the arrival and departure rates add up to one, there is a symmetry between arrivals and departures, which is evident in the figure.

\item We compute just the empirical death rate for the third phase, as we do not expect the birth rate to follow a linear trend during the third phase; see Figure~\ref{figure:phase3} for the resulting piecewise linear regression, where $R^2 = 0.9942$. We note that the departure rates are approximately piecewise linear following two linear trends: the left-hand linear trend in Figure~\ref{figure:phase3} is slightly upwards until the knot at occupancy number 32, and that to the right is a downwards trend. To better understand Figure~\ref{figure:phase3}, it is helpful to inspect it from right to left, which corresponds to the temporal sequence as people leave the space until it becomes empty. During the linear trend on the right, the departure rate increases as the occupancy decreases, which is not in accordance with the model as in (\ref{eq:death}). An obvious reason for this is that the imperative to leave the office increases towards the end of the day. During the linear trend on the left, which is slightly increasing, the departure rate is consistently high as its gets close to the end of the day.
\end{enumerate}

\section{Concluding Remarks}
\label{sec:conc}

We have provided a solution to the human dynamics problem of computing the population density of a social space, where agents enter and exit the space taking into account the proportion of the space currently occupied. As we have stressed throughout, estimating occupancy levels is one of the important tasks of building performance simulation.

\smallskip

We have proposed a model employing a birth-death process and shown that its solution follows a gamma distribution.
We extended the basic model to allow emigration and a special case of multiple types of agents.
We have validated the model using simulated data, and have given a proof of concept using a real-world data set of occupancy traces from a commercial office building. It would be desirable to further validate the model and study how it may be applied to various occupancy scenarios,
in addition to the typical workplace schedule as shown in Figure~\ref{figure:occ}.
However, this will necessitate obtaining access to additional suitable data sets.

\section*{Acknowledgements}

The authors would like to thank the reviewers for their constructive comments, which have helped us to improve the paper.


\newcommand{\etalchar}[1]{$^{#1}$}

\end{document}